\documentclass[preprint,showpacs,preprintnumbers,amsmath,amssymb]{revtex4}
\usepackage{graphicx}% Include figure files
\usepackage{dcolumn}% Align table columns on decimal point
\usepackage{bm}% bold math
\def\jpsi{{J/\psi}}

\def\be{\begin{equation}}
\def\ee{\end{equation}}
\def\bea{\begin{eqnarray}}
\def\eea{\end{eqnarray}}
\def\NO{\nonumber}

\def\dfrac{\displaystyle\frac}

\def\md{\mathrm{d}}

\def\a{\alpha}

\def\d{\delta}

\def\s{\sigma}

\begin{document}

%\preprint{APS/123-QED}

\title{Next-to-Leading-Order QCD corrections to $J/\psi(\Upsilon)+\gamma$ production at the LHC}% Force line breaks with \\

\author{Rong Li and Jian-Xiong Wang}%
% \email{Second.Author@institution.edu}
\address{
Institute of High Energy Physics, Chinese Academy of Sciences, P.O. Box 918(4),
Beijing, 100049, China.\\
Theoretical Physics Center for Science Facilities, CAS, Beijing, 100049, China.
}%
\date{\today}% It is always \today, today,
             %  but any date may be explicitly specified

\begin{abstract}
In this work, we calculate the next-to-leading-order (NLO) QCD
corrections to the process $p+p \to J/\psi + \gamma$ via the
color-singlet mechanism at the LHC. The results show that the
partial cross section ($p_t^{J/\psi}>$3GeV and
$|y^{J/\psi,\gamma}|<$3) is enhanced by a factor of about 2.0, and
the differential cross section can be enhanced by two orders of
magnitude in the large transverse momentum region of $J/\psi$.
Furthermore, the polarization of $J/\psi$ changes from transverse
polarization at leading-order to longitudinal polarization at NLO.
For the inclusive $J/\psi$ hadroproduction, it is known that the
color-octet contributions are one order of magnitude larger than the
color-singlet contribution, and the polarization distribution is
dominated by the color-octet behavior at NLO. In contrast, for
$J/\psi+\gamma$ production the color-singlet contribution is of the
same order as the color-octet contribution, and the polarization
distribution arises from both the color-singlet and color-octet.
Therefore, measurements of $J/\psi$ production associated with a
direct photon at the hadron collider could be an important
supplement to testify the theoretical framework treating with the
heavy quarkonium. In addition, an NLO QCD correction to
$\Upsilon+\gamma$ production at the LHC is also presented in this
paper.
\end{abstract}

%\begin{keyword}
%\PACS 12.38.Bx \sep 13.25.Gv \sep 13.60.Le
%\end{keyword}
% PACS, the Physics and Astronomy
                                   % Classification Scheme.
%\keywords{Suggested keywords}%Use showqkeys class option if keyword
                              %display desired
\maketitle
%\section{Introduction}
For a long time, the productions and decays of heavy quarkonium have
been an ideal laboratory to investigate quantum chromodynamics. The
large mass of the heavy quarks provides a large scale for production
and decay processes. This large scale makes the factorization of the
calculations for these processes available, and the hard part in the
calculations can be calculated perturbatively. Conventionally, the
color-singlet mechanism(CSM) \cite{Einhorn:1975ua} was used to
describe the production and decay of heavy quarkonium. However, the
CSM has encountered many difficulties in various
theoretical\cite{Barbieri:1976fp} and experimental
aspects\cite{Abe:1992ww}. In 1995, non-relativistic quantum
chromodynamics(NRQCD) was put forward\cite{Bodwin:1994jh}. By
including the contributions from the high Fock states, NRQCD can
overcome the theoretical difficulties\cite{Bodwin:1992ye}. At the
same time, the NRQCD predictions are consistent with experimental
data\cite{Braaten:1994kd}. Although NRQCD has had many successes,
there are also many problems in the production of heavy
quarkonium\cite{Brambilla:2004wf}.

In 1995, Kramer calculated the NLO QCD corrections to
inclusive $J/\psi$ photoproduction and found that the data from HERA
could be understood by just including the high order QCD corrections
in CSM \cite{Kramer:1995nb}. Recently, many studies
\cite{Zhang:2005cha,Gong:2007db} have shown that NLO QCD corrections
in CSM to $J/\psi$ related production processes at B-factory change
the theoretical predictions dramatically. Furthermore, at
the Tevatron, the NLO QCD corrections in
CSM enhance the total cross section by a factor of 2 and give large
values for the transverse momentum $p_t$ distribution of $J/\psi$,
especially in the large $p_t$
region\cite{Campbell:2007ws,Gong:2008sn}. The polarization
status of $J/\psi$ drastically changes from transverse-polarization
dominant at leading-order (LO) into longitudinal-polarization
dominant at NLO for $J/\psi$
hadroproduction\cite{Gong:2008sn,Gong:2008hk}. Artoisenet and his
collaborators suggested that the $p_t$ distribution of
$\Upsilon(1S)$ at the Tevatron can be interpreted in the CSM by
taking the NNLO real part into account \cite{Artoisenet:2008fc}. It
is quite clear that higher order QCD corrections play a very
important role in theoretical predictions for the production of
heavy quarkonium, and their effects should be carefully investigated in all the
relevant processes.

In the framework of CSM, the associated production of
$J/\psi+\gamma$ at a hadron collider was first proposed as a good
channel to investigate the gluon distribution in the proton with a
relatively clean signal\cite{Drees:1991ig}. Soon thereafter, it was
used to study the polarized gluon
distribution\cite{Doncheski:1993dj} and the production mechanism of
heavy quarkonium\cite{Kim:1994bm}. In reference \cite{Roy:1994vb},
the associated production of $J/\psi+\gamma$ at the Tevatron has
been considered in the CSM at the LO, and the results show that the
contribution from gluon fusion process is dominant over that from
the fragmentation processes. Kim investigated the contribution of
color-octet processes to the hadroproduction of $J/\psi+\gamma$ and
found that the color-octet contributions are dominant in the large
$p_t$ region\cite{Kim:1996bb}. The contribution from fragmentation
processes is smaller than the fusion contribution within the NRQCD
framework at the LHC\cite{Mathews:1999ye}. In reference
\cite{Kniehl:2002wd}, the authors obtained a theoretical prediction
for $J/\psi+\gamma$ hadroproduction at the hadron collider, and the
numerical results show that the transverse momentum distribution of
$J/\psi$ production is smaller in the CSM than that in the
color-octet mechanism (COM) in the large $p_t$ region at LO. To
further study the effect of the NLO QCD corrections on heavy
quarkonium hadroproduction, in this paper we calculate the NLO QCD
corrections to $J/\psi+\gamma$ hadroproduction at the LHC and
present theoretical predictions for the $p_t$ distribution of the
production and polarization for $J/\psi$.

We employ the automated Feynman Diagram Calculation package (FDC) to
perform the analytic evaluation of all the processes.
FDC is a powerful program based on the LISP language designed to
automate the NLO calculation and was initially
developed by Wang\cite{Wang:2004du}. Its one-loop part was recently completed by
Wang and Gong~\cite{fdc:2008}. It has recently been successfully
applied to several quarkonium production processes
\cite{Gong:2007db,Gong:2008sn,Gong:2008hk,Gong:2008ft,Gong:2008ue}.

%\section{}
For the process $p+p\to J/\psi+\gamma$ at LO, only the gluon fusion
process $g+g\to J/\psi+\gamma$ contributes at the partonic level with six Feynman diagrams,
which is similar to that of $g+g\to J/\psi+g$ in inclusive $J/\psi$ hadroproduction.
The LO total cross
section is obtained by convoluting the partonic cross section with
the parton distribution function (PDF) $G_g(x,\mu_f)$ in the proton: \be
\s^B=\int \mathrm{d}x_1\mathrm{d}x_2
G_g(x_1,\mu_f)G_g(x_2,\mu_f)\hat{\s}^B , \ee where $\mu_f$ is the
factorization scale. In the following, $\hat{\s}$ represents the
corresponding partonic cross section.

When the cross section is expanded to NLO on $\alpha_s$, the
coupling constant of quantum chromodynamics, there are one virtual
correction and three real correction processes. They are listed as the
following: \bea
g+g \to J/\psi +\gamma, \label{eqn:v}  \\
g+g \to J/\psi + \gamma +g, \label{eqn:r1}  \\
q(\bar{q})+g \to J/\psi + \gamma + q(\bar{q}), \label{eqn:r2}  \\
q + \bar{q} \to J/\psi + \gamma +g \label{eqn:r3}. \eea In
calculating the virtual corrections in Eq.(\ref{eqn:v}), there are
111 Feynman diagrams at NLO. The ultraviolet (UV) and infrared
(IR) singularities are normalized and separated by using the
dimensional regularization. Furthermore, the same renormalization
scheme as in Ref.~\cite{Gong:2008sn,Gong:2008hk} is applied to
redefine the quark mass, coupling constant and the quark or gluon
fields, and then the renormalized amplitude without UV singularities
is obtained. The Coulomb singularity, which comes from the diagrams
with a virtual gluon connecting the quark and anti-quark pair in
$J/\psi$, is regulated by introducing a small relative velocity $v$
between the quark and anti-quark pair and is absorbed into the
redefinition of the wave function of $J/\psi$. Therefore, the
one-loop amplitude $M^V$ is free of UV and Coulomb singularities and
the virtual corrections to the NLO cross section are expressed as
\be \dfrac{\mathrm{d}\hat{\s}^{V}}{\mathrm{d}t} \propto
2\mathrm{Re}(M^BM^{V*}), \ee where $M^B$ is the amplitude at LO. It
is UV and Coulomb finite, but has IR singularities. To obtain an
infrared-safe cross section, it is needed to cancel the IR
singularities by adding the contributions from real processes at NLO.

For the real processes in Eqs.(\ref{eqn:r1}), (\ref{eqn:r2}) and
(\ref{eqn:r3}), there are IR singularities in the
phase space integration which are divided into soft and  collinear
singularities. The soft singularity is from a soft gluon emitting
from the initial gluons. The collinear singularity is from the
initial gluon (or quark) emitting a gluon which is nearly parallel
to the parent particle. It is easy to find that soft singularities
caused by emitting soft gluons from the charm quark-antiquark pair
in $\jpsi$ are canceled by each other. Using the standard two
cut-off slicing method\cite{Harris:2001sx}, we decompose the phase
space into three regions by introducing two small cutoffs $\d_s$ and
$\d_c$. Therefore, convoluting the parton level cross section with
the parton distribution function, the cross section of the real
processes is represented as \be
\s^R=\s^{H\overline{C}}+\s^S+\s^{HC}+\s^{HC}_{add}. \ee Here
$\s^{H\overline{C}}$ is the contribution from the hard noncollinear
part of the phase space that is IR finite, $\s^S$ is the soft
part and $\s^{HC}$ is the hard collinear part. After absorbing
the mass factorization parts into the PDFs, there is a finite term
$\s^{HC}_{add}$ remaining which is similar to that in Eq.(43) in
Ref.\cite{Gong:2008hk}. By adding all the contributions together, a
finite total cross section is obtained.
%%%%%%%%%%%%%%%%%%%%%%%%%%%%%%%%%%%%%%%%%%%%%%%%%%%%%%%%%%%%%%%%%%%%%
\begin{figure}
\center{
\includegraphics*[scale=0.4]{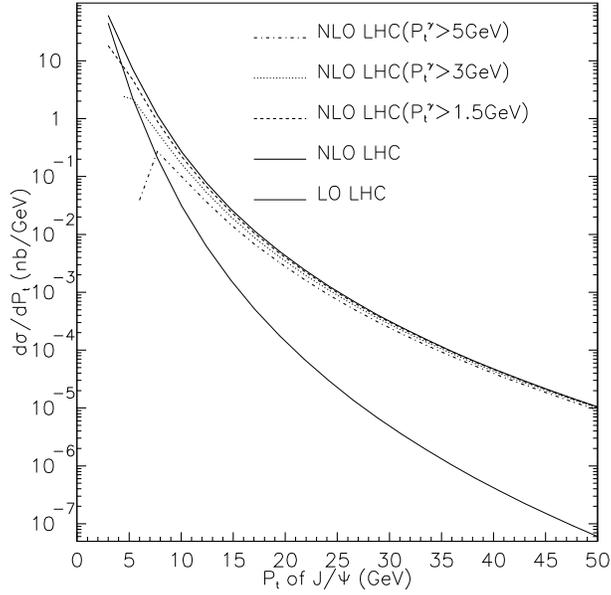}% Here is how to import EPS art
\caption {\label{fig:pt}Transverse momentum distribution of
$J/\psi$ production with $\mu_r=\mu_f=\sqrt{(2m_c)^2+p_t^2}$
at the LHC.}}
\end{figure}

The polarization of $J/\psi$ is an important physical measurement to
identify different production mechanisms. Therefore, the
polarization is calculated at NLO. Here we describe the decay of
$J/\psi$ in its rest frame with the 3-momentum direction of $J/\psi$
as the z-axis. The polarization parameter $\alpha$ is defined as \be
\alpha(p_t)=\frac{{\md\s_T}/{\md p_t}-2 {\md\s_L}/{\md
p_t}}{{\md\s_T}/{\md p_t}+2 {\md\s_L}/{\md p_t}}, \ee where $\s_T$
and $\s_L$ are the cross sections of transversely and longitudinally
polarized $J/\psi$ respectively. $\a=-1$ corresponds to fully
longitudinal polarization and $\a=1$ to fully transverse
polarization.

To check the gauge invariance
for each subprocess, the polarization vectors are replaced by the
corresponding momentum in the numerical calculation and the gauge invariance is
obviously observed at double precision level.
Since the two phase space cutoffs are chosen to handle the IR singularities of the real processes,
we varied the cutoffs by a few orders of
magnitude to numerically check the independence
of the results from the cutoffs and obtained consistent
results within the error tolerance.

%\section{numerical results}
In the numerical calculation, we use $\alpha$=1/137, $m_c$=1.5GeV,
and $M_{J/\psi}=2m_c$. From the leptonic width of $J/\psi$, the wave
function at the origin is extracted by
\bea
\Gamma_{ee}=(1-\frac{16\alpha_s}{3\pi})\frac{16\pi\alpha^2e_c^2}{M_{J/\psi}^2}|R(0)|^2.\label{eqn:ee}
\eea
We obtain $|R_{J/\psi}(0)|^2$=0.944GeV$^3$ at $\alpha_s=\alpha_s(M_{J/\psi})=0.26$, $\Gamma_{ee}=5.55$keV. The Cteq6L1 and
Cteq6M\cite{Pumplin:2002vw} are used in the calculations at LO and
NLO respectively, with the corresponding $\alpha_s$ running formula
in Cteq6 being used. The renormalization scale $\mu_r$ and the
factorization scale $\mu_f$ are set to
$\mu_r=\mu_f=\sqrt{(2m_c)^2+p_t^2}$. The center-of-mass energy at
the LHC is 14TeV and the rapidity cuts for both $J/\psi(\Upsilon)$
and the direct photon are chosen as $|y|<3$. Because the
perturbative expansion does not work well in the small transverse
momentum region and the large rapidity region, we impose $p_t$ and
$y$ cuts on the numerical calculation. In these processes, the
direct photon has nothing to do with the infrared singular
structure, so it can be identified with any experimentally
acceptable kinematic cuts. Our numerical calculation is obviously
consistent with this expectation. Of course, the cut conditions
imposed on the direct photon will affect the numerical results. In
the following, we fix the rapidity cut of the photon as $|y|<3$ and
vary its $p_t$ cuts to calculate the numerical results.

By replacing the corresponding parameters as
\bea
m_c  \leftrightarrow  m_b,~~~~M_{\jpsi}  \leftrightarrow  M_{\Upsilon},~~~~ e_c \leftrightarrow e_b \NO \\
R_s(0)^{J/\psi}\leftrightarrow R_s(0)^{\Upsilon},~~~~~n_f=3 \leftrightarrow n_f=4,
\eea
the results of $\Upsilon$ can be obtained. For the bottom quark mass and the 
wave function of $\Upsilon$ at the origin, $m_b$=4.75GeV and $|R_{\Upsilon}(0)|^2$=7.74GeV$^3$ are used. 
We use the Eq.(\ref{eqn:ee}) to calculate the $|R_{\Upsilon}(0)|^2$ with $\alpha_s(M_{\Upsilon})=0.18$ and $\Gamma_{\Upsilon\to ee}=1.34$keV.

\begin{figure}
\center{
\includegraphics*[scale=0.4]{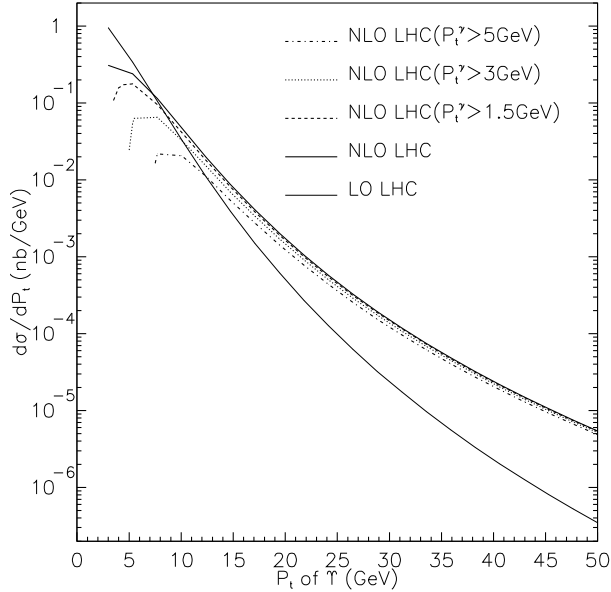}% Here is how to import EPS art
\caption {\label{fig:ptu}Transverse momentum distribution of
$\Upsilon$ production with $\mu_r=\mu_f=\sqrt{(2m_b)^2+p_t^2}$
at the LHC.}}
\end{figure}
%%%%%%%%%%%%%%%%%%%%%%%%%%%%%%%%%%%%%%%%%%%%%%%%%%%%%%%%%%%%%%%%%%%%%%%%%%%%%%%%%%%%%%%%%
The $p_t$ distribution of $J/\psi$ with different $p_t$ cut
conditions on the associated photon is shown in Fig.~\ref{fig:pt}.
It is noteworthy that there are two $p_t$ cut conditions, of which one is the
$J/\psi$ $p_t$ cut condition and the other is the associated photon
$p_t$ cut condition since both $J/\psi$ and the associated photon
have to be measured experimentally. The NLO numerical results of the
$p_t$ distribution of $J/\psi$ are larger than the LO results by
$1\sim 2$ orders of magnitude in the large $p_t$ region, and
different cut conditions only affect the results in the region
near the minimum endpoint in $p_t$. In the case of inclusive
$J/\psi$ hadroproduction at NLO in COM~\cite{Gong:2008ft}, the
results show that the NLO QCD corrections change the results by about 10
percent. Therefore, we can expect that the NLO QCD corrections will modify
the results of $J/\psi+\gamma$ at NLO in COM slightly, since the process
is quite similar to inclusive $J/\psi$ hadroproduction at
NLO in COM. Fig.~\ref{fig:ptu} shows that the NLO QCD corrections
also change the $p_t$ distribution of $\Upsilon$, but with less
enhancement than that of $J/\psi$.
%%%%%%%%%%%%%%%%%%%%%%%%%%%%%%%%%%%%%%%%%%%%%%%%%%%%%%%%%%%%%%%%%%%%%
\begin{figure}
\center{
\includegraphics*[scale=0.4]{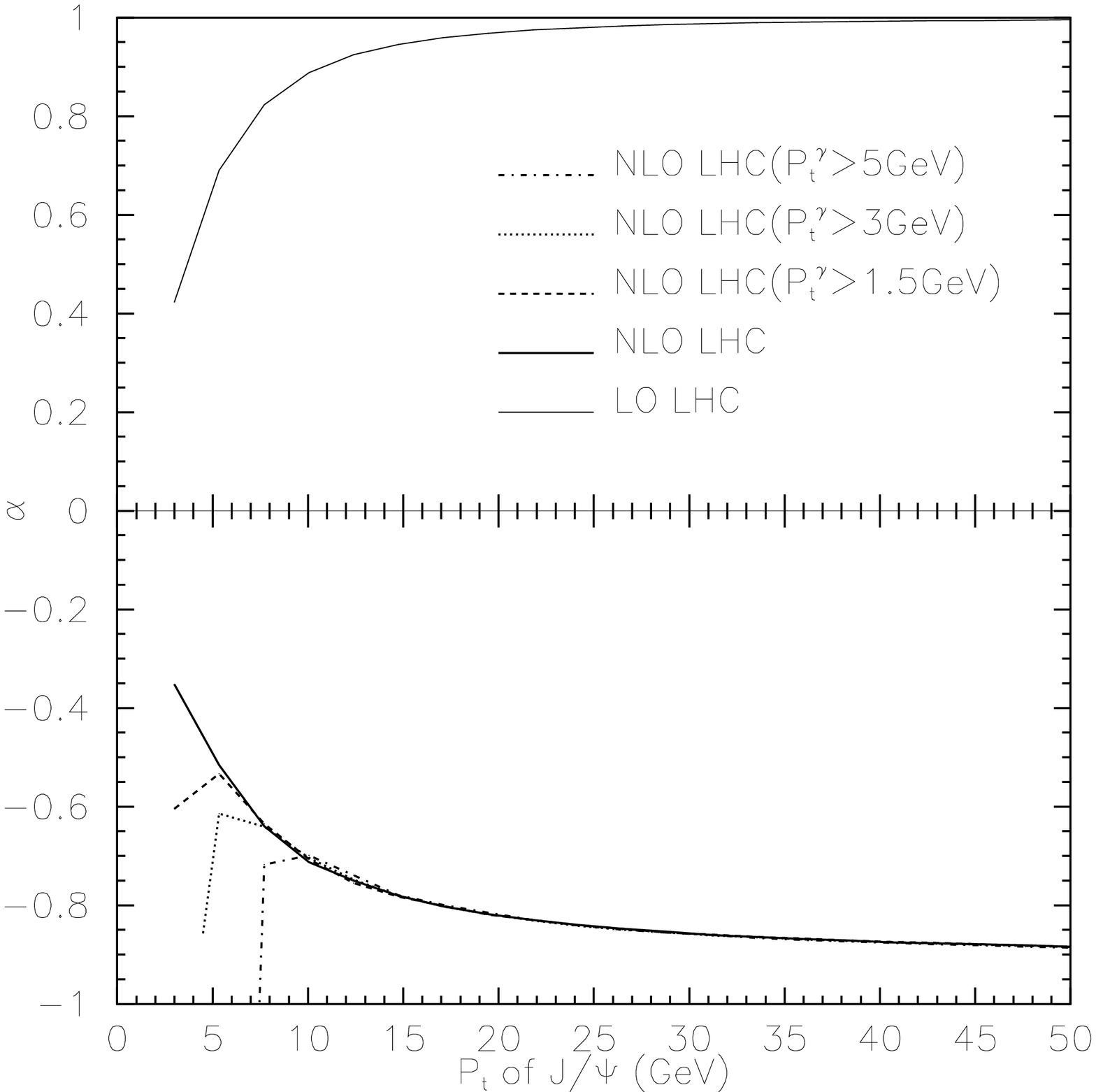}% Here is how to import EPS art
\caption {\label{fig:ptl}Transverse momentum dependence of
$J/\psi$ polarization with $\mu_r=\mu_f=\sqrt{(2m_c)^2+p_t^2}$ at
the LHC.}}
\end{figure}

\begin{figure}
\center{
\includegraphics*[scale=0.4]{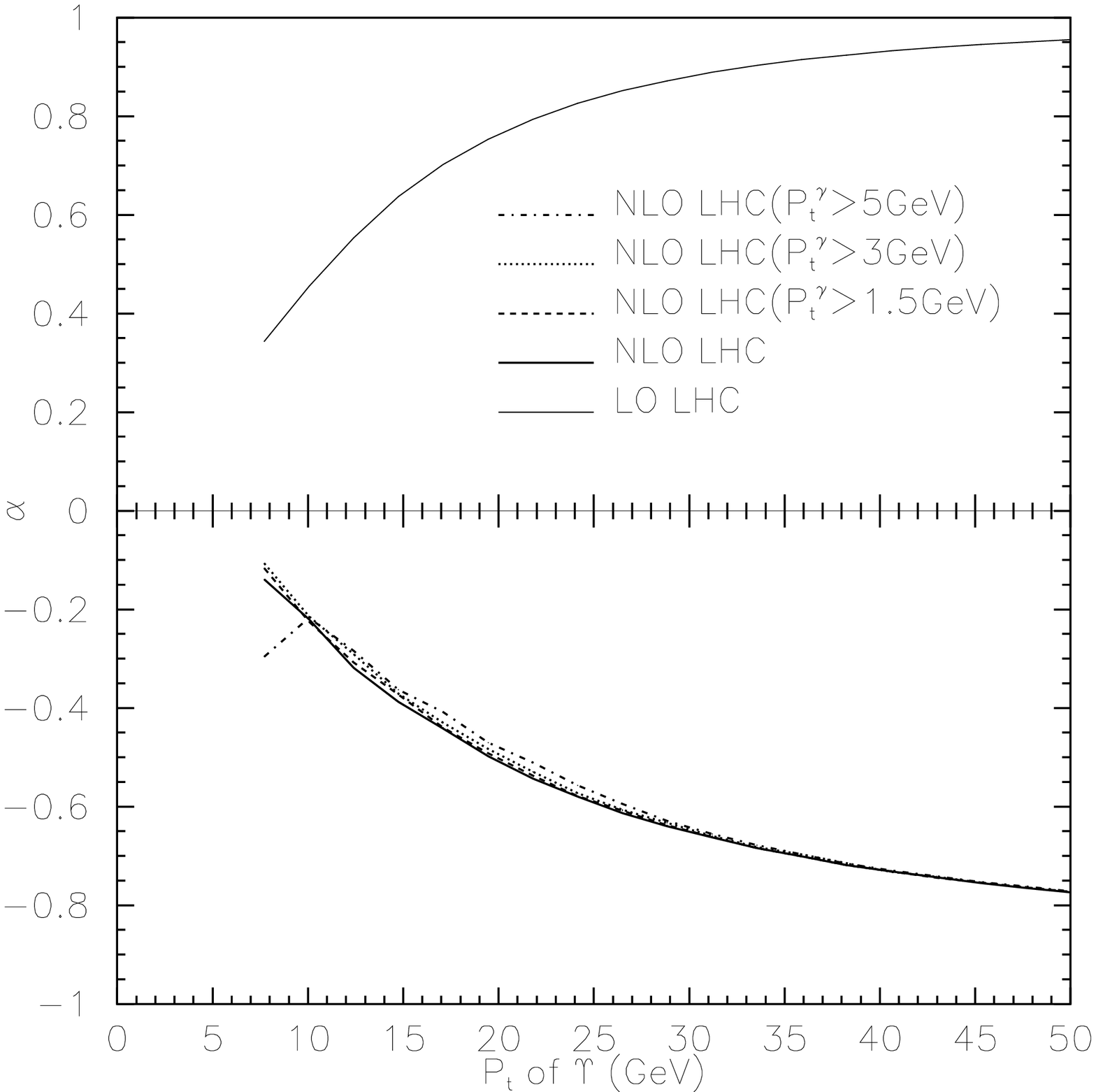}% Here is how to import EPS art^M
\caption {\label{fig:ptlu}Transverse momentum dependence of
$\Upsilon$ polarization with $\mu_r=\mu_f=\sqrt{(2m_b)^2+p_t^2}$ at
the LHC.}}
\end{figure}
%%%%%%%%%%%%%%%%%%%%%%%%%%%%%%%%%%%%%%%%%%%%%%%%%%%%%%%%%%%%%%%%%%%%%%%%%%%%%%%%%%%%%%%%%%%%
The theoretical predictions of the $J/\psi$ polarization at NLO are
presented in Fig.~\ref{fig:ptl} with different cut conditions. They
are similar to the result of $J/\psi$ inclusive hadroproduction. The
$J/\psi$ polarization drastically changes from the more transverse
polarization at LO to a more longitudinal polarization at NLO.
Also, the polarization parameter $\a$ of $J/\psi$ at NLO becomes
closer to -0.9 as $p_t$ increases. In the figures, the different
curves are plotted with different $p_t$ cut conditions. It is
clear that the $p_t$ cut condition for the associated photon has a greater
effect near the minimum endpoint of $J/\psi$ $p_t$ and a smaller
effect in the large $p_t$ region. From Fig.~\ref{fig:ptlu}, the
polarization of $\Upsilon$ also changes from transverse-polarization
dominant to longitudinal-polarization dominant, and $\a$ at NLO
becomes closer to -0.8 as $p_t$ increase. In the small $p_t$ region,
the convergence of the numerical calculation becomes worse, and it
requires too much CPU time to improve the precision of the
calculations. At the same time, the denominator of $\a$ changes its
sign in the neighborhood of a $p_t$ point in the small $p_t$ region,
meaning that the value of $\a$ undergoes a vast change around that point. The
small value of the denominator also amplifies the error in the
calculation, indicating that the convergence of the perturbative
expansion is bad and that higher order contributions are important in the
small $p_t$ region, especially for the $p_t$ distribution of the
polarization parameter $\a$. Therefore, the small $p_t$ region is further
discarded in our presentation of $\a$.
%%%%%%%%%%%%%%%%%%%%%%%%%%%%%%%%%%%%%%%%%%%%%%%%%%%%%%%%%%%%%%%%%%%%%%
\begin{figure}
\center{
\includegraphics*[scale=0.4]{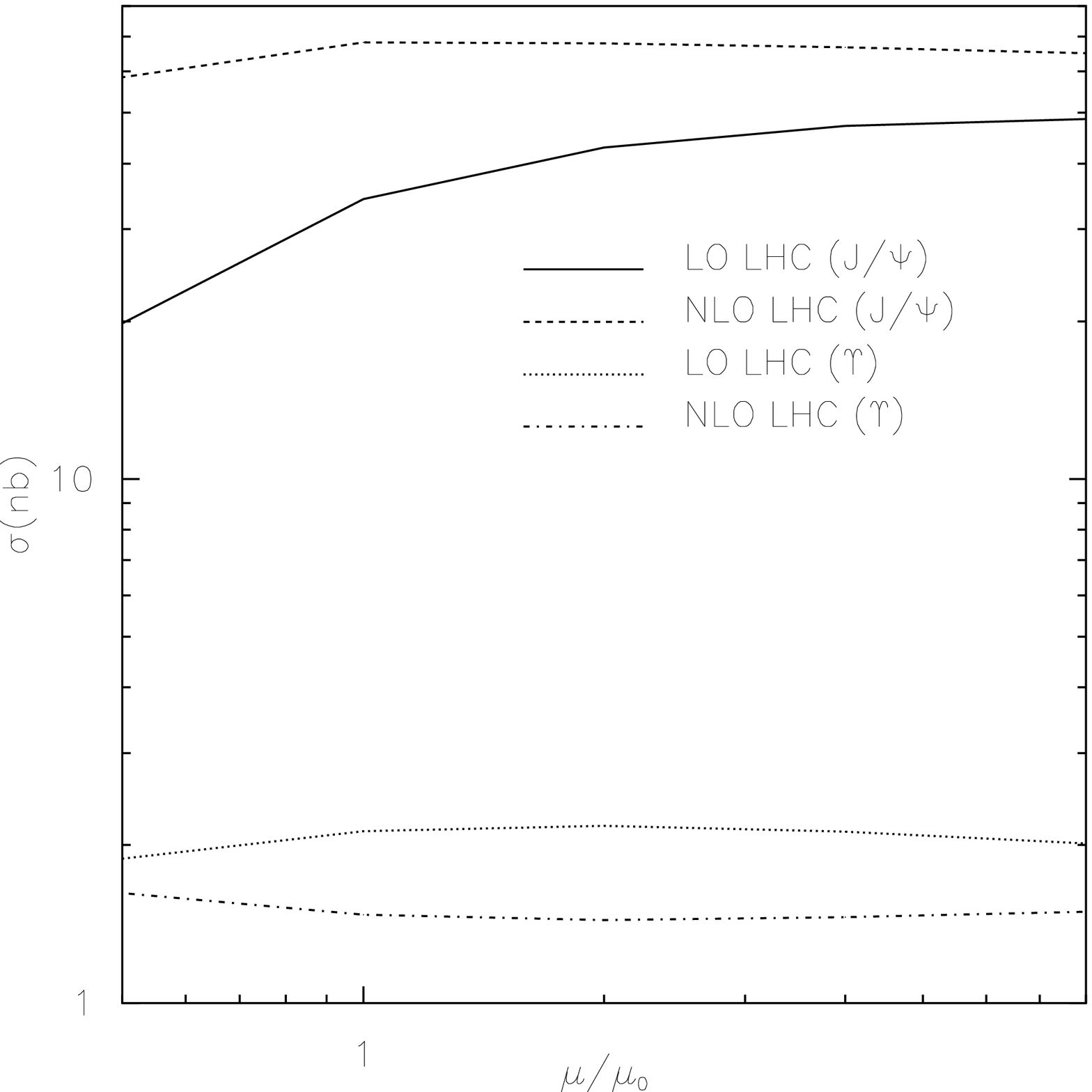}% Here is how to import EPS art
\caption {\label{fig:nocutto} The $\mu$ dependence of the partial
cross section ($p_t^{J/\psi(\Upsilon)}>$3GeV) with $\mu_0=\sqrt{(M_{J/\psi(\Upsilon)})^2+p_t^2}$. }}
\end{figure}
%%%%%%%%%%%%%%%%%%%%%%%%%%%%%%%%%%%%%%%%%%%%%%%%%%%%%%%%%%%%%%%%%%%%%%%%%%%%%%%%%%%%%%%%%%%%%

In Fig.~\ref{fig:nocutto}, the $\mu$ dependence of the partial cross
section ($p_t^{J/\psi(\Upsilon)}>$3GeV,$|y|<3$) for
$J/\psi(\Upsilon)+\gamma+X$ production at LO and NLO are shown
with our default choice, $\mu_r=\mu_f=\mu$. In other words, the factorization
scale equals to the renormalization scale, and $\mu$ ranges from
$\mu_0/2$ to $8\mu_0$. There is no $p_t$ cut for the associated
photon. It is obvious that QCD corrections reduce the $\mu$
dependence of the partial cross section for $J/\psi$ production and
moderate the $\mu$ dependence for $\Upsilon$ production.
%%%%%%%%%%%%%%%%%%%%%%%%%%%%%%%%%%%%%%%%%%%%%%%%%%%%%%%%%%%%%%%%%%%%%%

\begin{table}
\caption[]{The partial cross section at LO and NLO with different cut
conditions for transverse momentum $p_t$ but fixed rapidity cut condition $|y|<3$ for both $J/\psi(\Upsilon)$ and the direct photon. (units: $p_t$(GeV), $\sigma$(nb))} \label{tab:sigma}
\begin{center}
\renewcommand{\arraystretch}{1.5}
\[
\begin{array}{|c|c|c|c|c|c|c|c|c|}
\hline\hline
 p_t^{\gamma}& p_t^{J/\psi} & \sigma_{LO}^{J/\psi} &
 \sigma_{NLO}^{J/\psi} & K_{J/\psi} & p_t^{\Upsilon} & \sigma_{LO}^{\Upsilon} &
 \sigma_{NLO}^{\Upsilon} & K_{\Upsilon}\\ \hline
 >0.0& >3.0 & 34 & 68 & 2.0 & >3.0 & 2.1 & 1.5 & 0.71 \\ \hline
 >1.5& >3.0 & 34 & 34 & 1.0 & >3.5 & 1.7 & 0.94 & 0.55 \\ \hline
 >3.0& >4.5 & 5.5 & 6.0 & 1.1 & >5.0 & 0.82 & 0.46 & 0.56 \\  \hline
 >5.0& >6.0 & 1.2 & 1.2 & 1.0 & >7.5 & 0.24 & 0.16 & 0.67 \\ \hline \hline
\end{array}
\]
\renewcommand{\arraystretch}{1.0}
\end{center}
\end{table}
%%%%%%%%%%%%%%%%%%%%%%%%%%%%%%%%%%%%%%%%%%%%%%%%%%%%%%%%%%%%%%%%%%%%%%%%%%%%%%%%%%%%%%%%%%%%%
%%%%%%%%%%%%%%%%%%%%%%%%%%%%%%%%%%%%%%%%%%%%%%%%%%%%%%%%%%%%%%%%%%%%%%%%%%
\begin{figure}
\center{
\includegraphics*[scale=0.4]{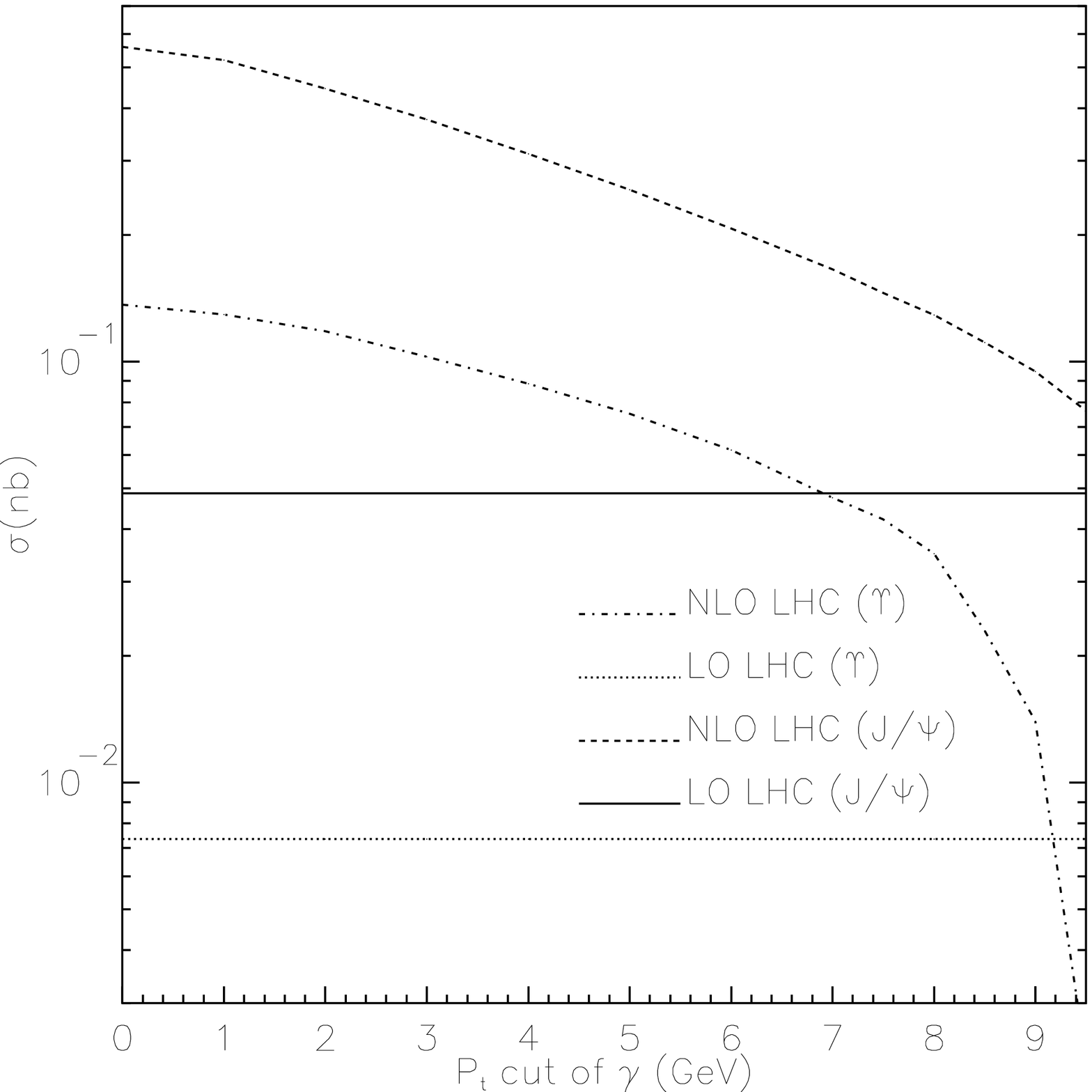}% Here is how to import EPS art
\caption {\label{fig:cutva}The partial cross section of the $J/\psi(\Upsilon)$
production associated with a direct photon as a function of $p_t$
cuts of the photon at the LHC. Here we fix the low bound of the $J/\psi(\Upsilon)$'s $p_t$ at 10GeV}}
\end{figure}
%%%%%%%%%%%%%%%%%%%%%%%%%%%%%%%%%%%%%%%%%%%%%%%%%%%%%%%%%%%%%%%%%%%%%%%%%%%%%%%%%%%%%%%%%%%%%%
To illustrate the influence of the cut conditions on the partial cross
section, we present LO and NLO partial cross sections for
$J/\psi(\Upsilon)+\gamma+X$ production at the LHC with different cut
conditions in Table ~\ref{tab:sigma}. In Fig.~\ref{fig:cutva}, we
plot the dependence of the partial cross section on the $p_t$ cut of
the direct photon with the $p_t$ cut of $J/\psi(\Upsilon)$ fixed at
10GeV. The plots show that the contributions from small $p_t$ photon
are large and the partial cross section decreases as the $p_t$ cut on
the photon increases. For the production of $\Upsilon$, when the
$p_t$ cut on the photon becomes too close to the $p_t$ cut of
$\Upsilon$, the partial cross section decreases rapidly and the
influence of the error becomes severe. In this case, too many
contributions from the real processes are cut off and the
perturbative calculation becomes worse. Therefore, the partial cross
section cannot be taken seriously when the $p_t$ cut condition of
the photon approaches the $p_t$ cut condition of $\Upsilon$.

In summary, we have calculated NLO QCD corrections to the
production of $J/\psi(\Upsilon)$ associated with a direct photon at
the LHC in CSM. For $J/\psi$ production, the partial cross
section is enhanced by a factor of about 2.0 with $J/\psi$
transverse momentum cut $p_t>$3GeV and rapidity cut $|y|<3$ for
both $J/\psi$ and the direct photon. The transverse momentum
distribution of $J/\psi$ at NLO is enhanced by $1\sim 2$ orders of
magnitude over LO calculations as $p_t$ becomes larger. The $J/\psi$
polarization is calculated, and the results show that the $J/\psi$
polarization drastically changes from transverse-polarization
dominant at LO to longitudinal-polarization dominant at NLO. The
situation is quite similar to the case of NLO QCD corrections to
the inclusive $J/\psi$ hadroproduction. It can be seen that NLO
results on $J/\psi+\gamma$ hadroproduction in CSM is of the same
order of magnitude as LO results in
COM\cite{Kim:1996bb,Kniehl:2002wd}. As a reasonable estimate
from the experience of NLO QCD correction calculation to
inclusive $J/\psi$ hadroproduction in COM\cite{Gong:2008ft}, the NLO
$p_t$ distribution for $J/\psi+\gamma$ hadroproduction in COM will
be changed slightly and the polarization will remain almost the
same. Therefore, the results of $J/\psi$ production associated with
a direct photon from the CSM and COM are of the same order of
magnitude at NLO. Therefore, the $J/\psi$ polarization measurement
would be able to distinguish the contributions from the CSM and COM since
the theoretical predictions of $J/\psi$ polarization are obviously
different for the CSM and COM. In contrast to the case of
$J/\psi$ inclusive hadroproduction, the NLO result of CSM is about
one order of magnitude smaller than the NLO result of COM, and so
there is no way for the CSM part to play an important role in the final
$J/\psi$ polarization distribution.

To measure the production of $J/\psi(\Upsilon)$ associated with a
direct photon at the LHC, the direct photon must be identified. To
cut the enormous background of photons emitted in jet hadronization,
photons from jets are attributed to the jets and a direct photon,
which is isolated from any jets, should be identified. Therefore,
being quite different from $J/\psi(\Upsilon)$ inclusive
hadroproduction, the hadroproduction of $J/\psi(\Upsilon)$
associated with a direct photon will be an important process to
investigate the production mechanism of heavy quarkonium and impose
new constraints on the contribution of COM to heavy quarkonium
hadroproduction.

We thank Gong Bin for helpful discussions. This work is supported by the National
Natural Science Foundation of
China (No.~10775141) and by the Chinese Academy of Sciences under
Project No. KJCX3-SYW-N2.

\bibliography{gg-Jpsigamma}% Produces the bibliography via BibTeX.
\end{document}